\documentclass[%
 reprint,
superscriptaddress,
 amsmath,amssymb,
 aps,
]{revtex4-2}

\usepackage{graphicx}
\usepackage{dcolumn}
\usepackage{bm}
\usepackage{xcolor}

\begin{document}

\preprint{APS/123-QED}

\title{Local composition fluctuations act as precursors for crystal nucleation in polydisperse hard spheres}


\author{Marjolein de Jager}
\affiliation{Soft Condensed Matter and Biophysics, Debye Institute for Nanomaterials Science, Utrecht University, 3584 CC Utrecht, Netherlands}
\author{Antoine Castagnède}
\affiliation{Universit\'e Paris-Saclay, CNRS, Laboratoire de Physique des Solides, 91405 Orsay, France}
\author{Frank Smallenburg}
\affiliation{Universit\'e Paris-Saclay, CNRS, Laboratoire de Physique des Solides, 91405 Orsay, France}
\author{Laura Filion}
\email{l.c.filion@uu.nl}
\affiliation{Soft Condensed Matter and Biophysics, Debye Institute for Nanomaterials Science, Utrecht University, 3584 CC Utrecht, Netherlands}

\date{\today}

\begin{abstract}
We revisit the effect of polydispersity on the crystal nucleation of hard spheres. Using event-driven molecular dynamics simulations, we obtain the nucleation rate as a function of the supersaturation for a range of polydispersities, and demonstrate that the nucleation rate of polydisperse hard spheres deviates from the trend of monodisperse hard spheres, even when mapped to the effective packing fraction.
Furthermore, we show that nucleation tends to originate in regions with on average more larger-sized particles, indicating that such regions act as precursors for nucleation in systems of polydisperse hard spheres.
\end{abstract}

\maketitle


\newcommand{\comment}[1]{{\color{red}{\bf #1}}}


Hard spheres serve as the archetypical model system for studying colloidal self-assembly, providing fundamental insights into a wide range of physical phenomena \cite{royall2024colloidal}. In their simplest form, monodisperse hard spheres have been instrumental in advancing our understanding of, e.g., 
phase transitions \cite{hoover1968melting,frenkel1984new,pusey1986phase,auer2001prediction,gasser2001real,filion2010crystal,thorneywork2017two,richard2018crystallizationI}, glassy behavior \cite{gordon1976hard,pusey1987observation,speedy1998hard,pusey2009hard,parisi2010mean}, and defects \cite{bowles1994cavities,bennett1971studies,pusey1989structure,pronk1999can,pronk2001point,van2017diffusion}.
In particular, the nucleation behavior of monodisperse hard spheres has drawn significant attention, especially due to the persistent discrepancy between computationally predicted nucleation rates and experimental observations \cite{auer2001prediction,filion2011simulation,russo2013interplay,radu2014solvent,wood2018coupling,fiorucci2020effect,dejager2022crystal,wohler2022hard,royall2024colloidal, kurten2025free, schope2007effect}. 
Moreover, significant effort has also been dedicated to understanding the nucleation mechanism itself, including the role of precursors \cite{schilling2010precursor,lu2015experimental,taffs2016role,berryman2016early,dejager2023search} and polymorph selection \cite{russo2012microscopic,gispen2023crystal}.

Moving beyond monodisperse systems, introducing polydispersity into the hard-sphere model brings new complexity and intriguing possibilities. Polydispersity not only alters the existing phase boundaries \cite{mcrae1988freezing,bolhuis1996numerical,sollich2010crystalline,fasolo2003equilibrium,fasolo2004fractionation,castagnede2025freezing,zaccarelli2009crystallization}, it also opens the door to new stable crystal structures \cite{bommineni2019complex,bommineni2020spontaneous,lindquist2018communication,cabane2016hiding,bareigts2020packing}. One particularly interesting consequence of polydispersity is fractionation, where different particle sizes distribute unevenly across coexisting phases. 
A striking example of fractionation is its role in the formation of Laves phases in highly polydisperse systems \cite{lindquist2018communication,bommineni2019complex,bommineni2020spontaneous,cabane2016hiding}, where the particles that are selected to be part of the crystal form a clear bimodal distribution, with each peak corresponding to different lattice positions.
However, fractionation is not limited to such exotic phases; in moderately polydisperse systems, where the face-centered cubic (FCC) crystal remains the stable crystal structure, fractionation plays a role in the fluid--crystal coexistence. In such coexistences, the crystalline phase exhibits a narrower size distribution with on average more larger particles than the fluid \cite{kofke1999freezing,martin2003crystallization,fasolo2003equilibrium,fasolo2004fractionation,castagnede2025freezing}.
This raises the question of whether certain local fluctuations in particle composition could act as precursors for nucleation in polydisperse systems, a phenomenon that is impossible in monodisperse hard spheres.

In this Letter, we revisit the effect of polydispersity on the nucleation rate of hard spheres and report evidence for a precursor for the nucleation of polydisperse hard spheres. 
To this end, we use event-driven molecular dynamics (EDMD) simulations \cite{alder1959studies,smallenburg2022efficient} to simulate systems of monodisperse and polydisperse hard spheres. The spheres have equal mass $m$ and their polydispersity $\delta$ is set via a Gaussian distribution with average diameter $\bar{\sigma}$. 
We do not make use of a thermostat; hence, the kinetic energy of the system is fixed, which also fixes the temperature $T$.
The time unit of our simulations is then given by $\tau=\sqrt{m \bar{\sigma}^2/k_BT}$, where $k_B$ is the Boltzmann constant.

We consider polydispersities 0\% to 6\%, and perform 100 independent brute force simulations of $N=10^4$ particles for a range of fluid packing fractions at each polydispersity. We use an implementation of the Lubachevsky-Stillinger approach \cite{lubachevsky1990geometric} to quickly quench the system to the desired packing fraction, after which we run each simulation for a maximum of $t_\text{max}=10^6\tau$. Simulations are terminated prematurely if the system contains a nucleus that has reached a size of $800$ particles or more, well beyond the critical cluster size at the state points we study.

To determine the sizes of the nuclei in a system, we classify each particle as either fluid or crystal using the six-fold ten Wolde bonds \cite{tenwolde1996simulation}
\begin{equation}\label{eq:d6}
    d_6(i,j) = \frac{ \sum_m q_{6m}(i) q_{6m}^*(j) }{ \sqrt{ \left(\sum_m |q_{6m}(i)|^2 \right) \left(\sum_m |q_{6m}(j)|^2 \right) } },
\end{equation}
where $^*$ denotes the complex conjugate, $\sum_m$ represents the sum over $m\in[-6,6]$, and $q_{6m}$ are Steinhardt's six-fold bond-orientational order parameters \cite{steinhardt1983bond}. Particle $i$ is classified as crystal if it has at least seven neighboring particles $j$ with which it has a crystal-like bond, i.e. $d_6(i,j)>0.7$. To identify the nearest neighbors, we use the solid angle nearest neighbor (SANN) algorithm \cite{van2012parameter}.


\begin{figure*}[t!]
    \centering
     \includegraphics[width=\linewidth]{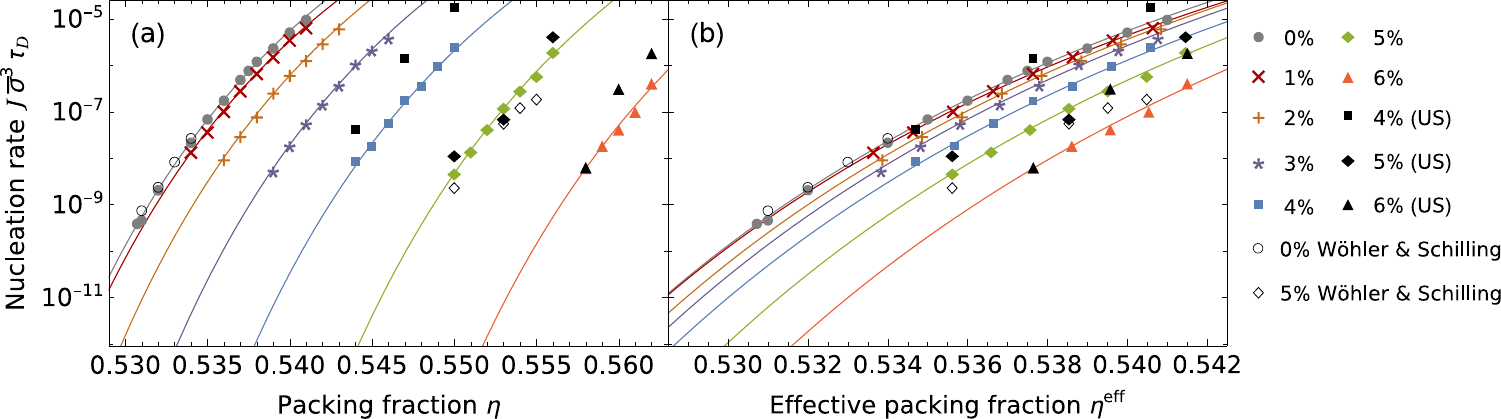}
    \caption{\label{fig:rate} 
    Nucleation rate as a function of (a) packing fraction and (b) effective packing fraction $\eta^\text{eff}=(\eta_F^0 / \eta_F)\eta$ for monodisperse and polydisperse hard spheres. The lines serve as guides to the eye, and indicate CNT-like fits (Eq. \ref{eq:fitrate}). 
    The solid black symbols indicate the rates obtained via umbrella sampling and the open symbols indicate the rates reported by W\"ohler \& Schilling \cite{wohler2022hard}.
    The nucleation rates are given in terms of the long-time diffusion time $\tau_D$.
    }
\end{figure*}

We obtain the nucleation rate from these brute force nucleation events via
\begin{equation}
    J = \frac{1}{\langle t_\text{nuc}\rangle V},
\end{equation}
where $V$ is the volume of the simulation box and $\langle t_\text{nuc}\rangle$ is the average nucleation time. Assuming that the nucleation times have a censored exponential distribution \cite{deemer1955estimation}, $\langle t_\text{nuc}\rangle$ is obtained via 
\begin{equation}
    \langle t_\text{nuc}\rangle = \frac{1}{k} \sum_{i=1}^n t_i.
\end{equation}
Here, $n$ indicates the total number of simulations, $k$ indicates the number of successful nucleation events, and $t_i$ indicates the time at which simulation $i$ starts its successful nucleation. 
In practice, we take for $t_i$ the last time instance at which a nucleus contains fewer than 100 particles. 
Note that, for simulations where no nucleation occurred, $t_i=t_\text{max}$. 
The resulting nucleation rates as a function of the fluid packing fraction $\eta$ are shown in Fig. \ref{fig:rate}a. We present the nucleation rates in terms of the long-time diffusion time $\tau_D=\bar{\sigma}^2/(6D_L)$, where $D_L$ is the long-time diffusion coefficient. We obtain the long-time diffusion coefficient directly from the simulations for the polydisperse systems, while for the monodisperse case we use the fitted expression by Erpenbeck and Wood \cite{erpenbeck1991self}.

Assuming that the nucleation rate follows classical nucleation theory (CNT), we can fit them using
\begin{equation}\label{eq:fitrate}
    J\bar{\sigma}^3\tau_D=a\exp\left[-b/(\eta-\eta_F)^2\right].
\end{equation}
Here, $\eta_F$ indicates the freezing packing fraction of the system and $a$ and $b$ are fit parameters. The parameter $a$ serves as the (dimensionless) kinetic prefactor of the nucleation rate and the term in the exponent represents the height of the nucleation barrier. To obtain this expression, we use that the height of the nucleation barrier is given by $16\pi\gamma^3/3(\rho_x|\Delta \mu|)^2$, where $\gamma$ is the interfacial free energy, $\rho_x$ is the density of the crystal phase, and $|\Delta \mu|$ is the chemical potential difference between the fluid and crystal phases \cite{kelton1991crystal}. For systems near the freezing point, we can assume that $\gamma$ is approximately constant and that $\rho_x|\Delta\mu|$ is linear in ($\eta-\eta_F)$.
The lines in Fig. \ref{fig:rate} represent the CNT-like fits through our results, and the values of their fit parameters are reported in the Supplementary Material.

We compare our results to the rates obtained by W\"ohler \& Schilling \cite{wohler2022hard} (open symbols in Fig. \ref{fig:rate}), and find good agreement for the monodisperse hard spheres. For the 5\% polydisperse hard spheres, we find that the rates differ by an approximate factor two. 
To further explore the nucleation rates, we additionally perform simulations using umbrella sampling (US) \cite{torrie1974monte,tenwolde1996simulation} for the systems of 4\%, 5\%, and 6\% polydispersity.
We employ a biasing potential in the form of a quadratic function of the size $n$ of the largest crystalline cluster found in the system, again identified using six-fold ten Wolde bonds:
\begin{equation}
    U_\textrm{bias} = \alpha (n-n_0)^2,
\end{equation}
where $n_0$ is the target cluster size for the simulation window considered.
Specifically, we use a spring constant $\alpha = 0.2 k_BT$ and increase the target cluster size by 10 units from one simulation window to another.
Simulations with $N = 10^4$ particles are run sequentially for each state point, starting from an equilibrated fluid configuration, and consisting of $10^6$ Monte-Carlo cycles including $4 \times 10^5$ equilibration cycles using swap moves \cite{kuchler2023understanding}.
The latter allow us to make sure that the formed crystalline cluster is well equilibrated with the surrounding fluid on the timescales we consider here. 
Kinetic prefactors to the nucleation rates are then evaluated by measuring the rate of fluctuations in cluster size at the top of the barrier. Specifically, we follow the methodology proposed in Refs. \cite{auer2001prediction, filion2010crystal}, using EDMD simulations starting at the top of the nucleation barriers.

The resulting rates are indicated by the solid black markers in Fig. \ref{fig:rate}, and we show the nucleation barriers in the Supplementary Material.
We find reasonable agreement between the brute force and US results, although the US rates appear generally faster, but are more noisy. Faster rates could be attributable to the fact that the US simulations sample equilibrium configurations, in which the size distribution of the crystalline clusters is adapted to each cluster size. As a result, these clusters are expected to be more stable than the kinetically formed clusters in the MD simulations, which do not have the chance to change their internal composition during the relatively short time span associated with crystal nucleation. Hence, we believe that the MD trajectories effectively follow a nucleation pathway with a (slightly) higher free-energy barrier. However, the noise in the US data makes it difficult to be certain on this point.

In order to compare the rates of the different systems (Fig. \ref{fig:rate}a), it is common practice (particularly in terms of experiments) to map each system onto the monodisperse hard-sphere system \cite{royall2024colloidal}. 
While there are several approaches for performing this mapping, the most common method is to align the freezing packing fractions of the systems.
The effective packing fraction of the system is then defined as 
\begin{equation} \label{eq:etaeff}
    \eta^\text{eff}= \frac{\eta_F^0 }{ \eta_F} \; \eta,
\end{equation}
where $\eta_F^0$ is the freezing packing fraction of monodisperse hard spheres, and $\eta$ and $\eta_F$ are, respectively, the fluid packing fraction and freezing packing fraction of the system under consideration. 
We use the freezing packing fractions reported in Ref. \cite{castagnede2025freezing}.
The resulting effective trends of the nucleation rate are shown in Fig. \ref{fig:rate}b.
Although this mapping results in a near agreement between the trends for the lower polydispersities, significant deviations from the trend of monodisperse hard spheres remain at higher polydispersities. These observations stand in contrast to previous claims in the literature that (low) polydispersity does not significantly affect nucleation rates when mapped to hard spheres
\cite{filion2010crystal,pusey2009hard}. 
However, care should be taken when comparing these claims to our results.
First, the mapping to the monodisperse systems is highly sensitive to errors in the determination of the freezing packing fraction. Small inaccuracies in $\eta_F$ can lead to significant differences in the mapped values of the effective packing fraction, thereby influencing the comparison of nucleation rates. 
Here, we use the recently published values from Ref. \cite{castagnede2025freezing}, based on accurate direct-coexistence simulations where the fluid follows the same Gaussian size distribution as we consider here.  Second, the remaining deviation between the two most extreme polydispersities in Fig. \ref{fig:rate}b (i.e. 0\% and 6\%) only corresponds to a shift in packing fraction of approximately $0.004$. This is a small shift in comparison to typical experimental errors in determining colloidal packing fractions: even with very careful real-space comparisons, it is difficult to get this error below $0.0025$ \cite{kurten2025free}.
Finally, previous studies exploring the effects of polydispersity \cite{auer2001prediction,auer2004numerical,filion2010crystal,pusey2009hard} often considered a broader range of supersaturations and examined larger-scale trends in the nucleation rate, as their primary focus was on the comparison with experimental results. In plots spanning a wider range of supersaturations and rates, the subtle deviations between the nucleation rates of polydisperse and monodisperse systems would be less visible.


\newcommand{\figwidthC}{0.45\linewidth}
\begin{figure*}[t!]
\centering
\begin{tabular}{ll}
     \includegraphics[width=\figwidthC]{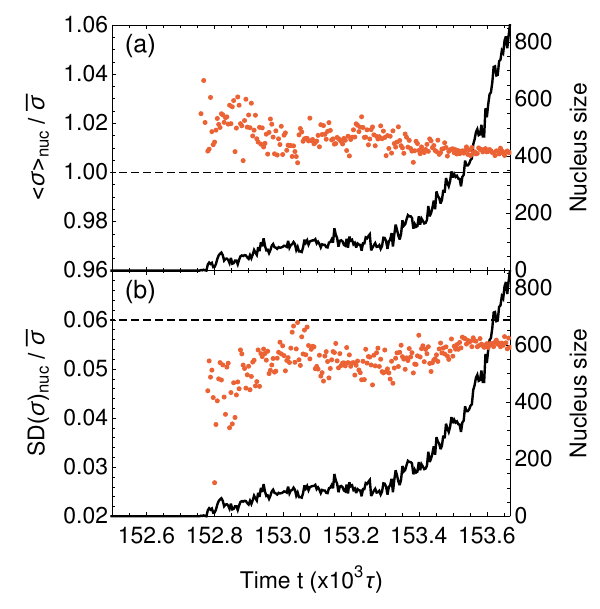} & \includegraphics[width=\figwidthC]{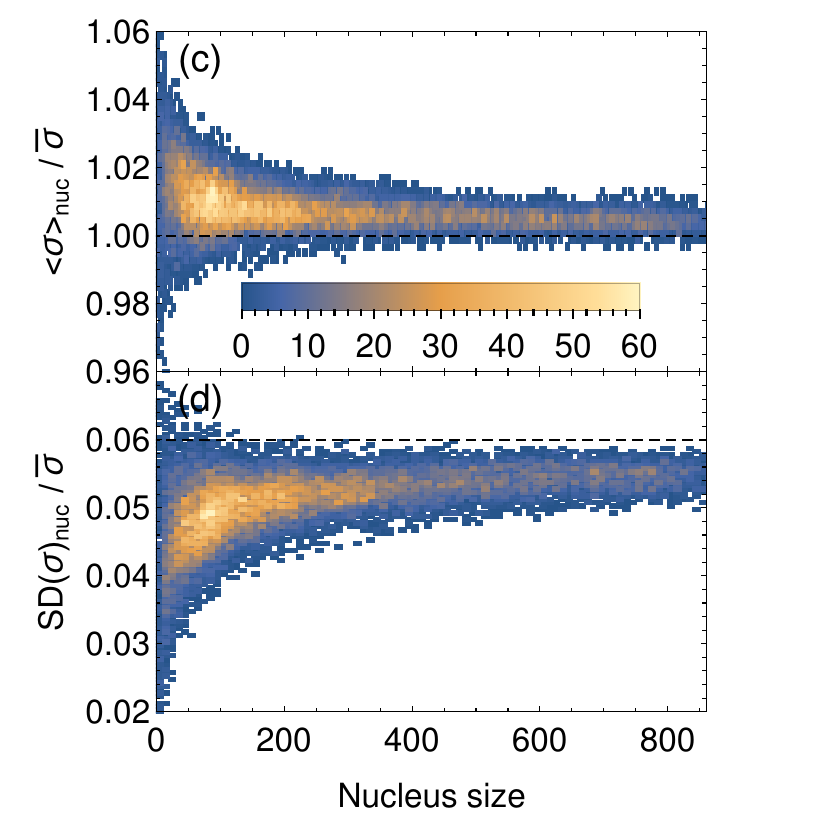} 
\end{tabular}
    \caption{\label{fig:compo} 
    (a) Average and (b) standard deviation of the particle size inside the crystal nucleus as a function of time for a typical nucleation event (6\% polydispersity at $\eta=0.562$). The black line indicates the nucleus size as a function of time. Note that the first $152.5\cdot10^3\tau$ of the simulation are not shown.
    In (c,d), we combine the data on all 100 nucleation events of the same system (6\% polydispersity at $\eta=0.562$) and show (c) the average particle size versus the nucleus size and (d) the standard deviation of the particle size versus the nucleus size. 
    The color gradient indicates the bin count.
    }
\end{figure*}

\begin{figure*}[t!]
    \centering
     \includegraphics[width=0.95\linewidth]{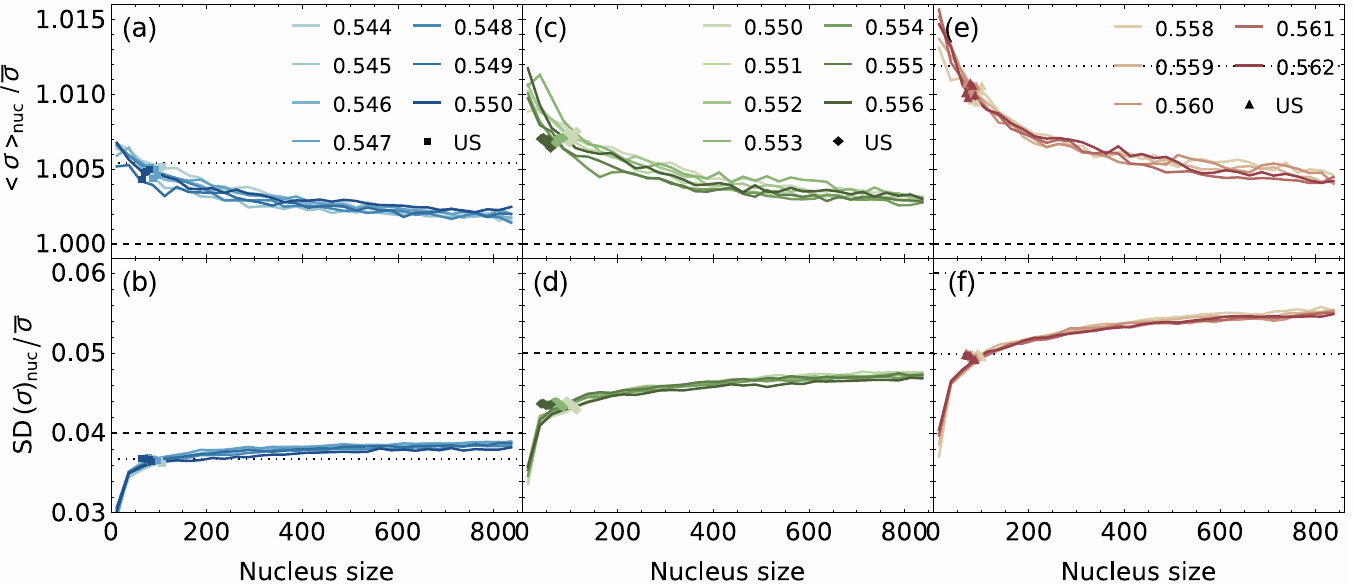}
    \caption{\label{fig:avcompo} 
    Average trends of the nucleus composition for (a,b) 4\%, (c,d) 5\%, and (e,f) 6\% polydispersity. These average trends are obtained from density histograms such as depicted in Fig. \ref{fig:compo}c,d.
    The different colored lines indicate different packing fractions of the system.
    Symbols correspond to critical nuclei from US simulations, and dotted lines are the composition of the equilibrium crystal phase obtained from direct coexistence simulations, taken from Ref. \cite{castagnede2025freezing}. Note that Ref. \cite{castagnede2025freezing} did not provide data for $5\%$ polydispersity.
    }
\end{figure*}


Next, we investigate the nucleation mechanism of polydisperse hard spheres. For this, we focus on the systems with 4\%, 5\%, and 6\% polydispersity.
As a starting point, we examine the particle size distribution of the successful nuclei. Figures \ref{fig:compo}a-b show, for a typical nucleation event of 6\% polydisperse hard spheres, the average and standard deviation of the particle size inside the crystal nucleus as a function of time. We clearly see that the particles inside the nucleus are both larger and less polydisperse: fractionation already occurs in the early crystal nucleus. In order to capture this behavior for all 100 nucleation events of this state point, we compute the density histogram of the average and standard deviation of the particle size inside the crystal nucleus versus the nucleus size. 
The resulting density histograms, shown in Figures \ref{fig:compo}c-d, confirm the same trends observed in Figures \ref{fig:compo}a and \ref{fig:compo}b: nuclei tend to have a narrower particle size distribution, shifted toward larger particle sizes in comparison to the total system.

These density histograms also allow us to extract the average trends for each system and state point studied, as shown in Fig. \ref{fig:avcompo}. From this figure, we draw three important observations.
First, the degree of supersaturation appears to have no significant influence on the average composition of the nuclei. 
Second, the smaller (precritical) nuclei---those containing around 100 particles or fewer---exhibit the strongest preference for larger particles. This suggests that once a nucleus enters the growth stage, its preference for larger particles diminishes, and it absorbs any neighboring fluid particle, regardless of size. In the Supplementary Material, we further demonstrate this point by examining the radial composition of the nuclei. 
Third, increasing the system's polydispersity enhances the preference of the nuclei for larger-sized particles. 

We also report in Fig. \ref{fig:avcompo} the composition of crystalline clusters at the top of the nucleation barrier as obtained from US simulations (plain symbols).
Again, we find good agreement with the brute force results.
Though these clusters are better equilibrated than those in the brute force simulations, this does not appear to strongly affect their composition. Interestingly, the average particle size and polydispersity in these critical nuclei are close to those of the (shadow) crystal that coexists with the fluid at its equilibrium freezing point (dotted lines in Fig. \ref{fig:avcompo}, taken from Ref. \cite{castagnede2025freezing}), albeit with a slightly smaller average particle size. 
It is interesting to compare these results to experiments on slightly charged colloidal spheres. Consistent with our observations here, Kurita \textit{et al.} found that for these charged particles, the crystal preferentially absorbed particles with a more monodisperse size distribution, especially during early stages of growth  \cite{kurita2012measuring}.



\begin{figure*}[t!]
    \centering
     \includegraphics[width=0.95\linewidth]{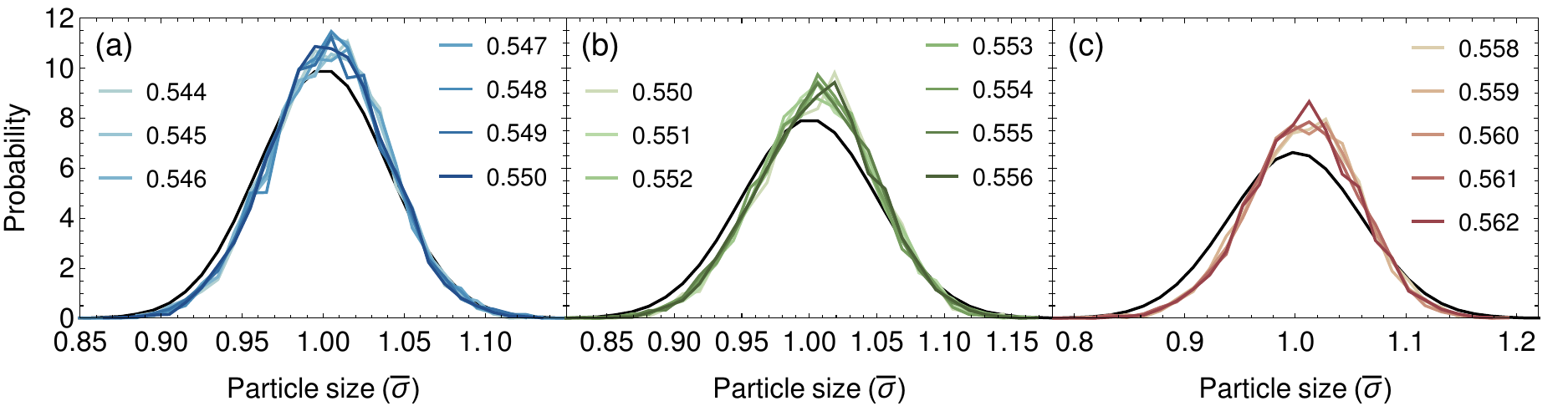} 
    \caption{\label{fig:onset} 
    Size distribution of the fluid region from which the nuclei originate for (a) 4\%, (b) 5\%, and (c) 6\% polydispersity. 
    The different colored lines indicate different packing fractions of the system, and the black line indicates the size distribution of the entire system. 
    Assuming that the supersaturation has no significant effect on the distribution, the means of the distributions are (a) $1.0032\bar{\sigma}$, (b) $1.0053\bar{\sigma}$, and (c) $1.0071\bar{\sigma}$, and the widths (i.e. standard deviations) are (a) $0.0367\bar{\sigma}$, (b) $0.0444\bar{\sigma}$, and (c) $0.0505\bar{\sigma}$. 
    }
\end{figure*}

Lastly, to investigate the onset of nucleation more closely, we measure the composition of the fluid in the region where the nucleus forms \cite{berryman2016early, dejager2023search}. 
For each nucleation event, this region is defined as a sphere of radius $2.2\bar{\sigma}$ centered at the nucleus' center of mass when it reaches a size of 100 particles. This region typically contains around 40-50 particles. We then identify the last instance for which no crystal particles are present in this region and record the sizes of the particles inside. By doing this for all successful nucleation events of a given state point, we can compute the local particle size distribution of the fluid just before a nucleation event begins. 
Figure \ref{fig:onset} shows the resulting size distributions for all systems and state points studied. These distributions are clearly narrower and shifted toward larger particle sizes compared to the distribution of the total system, aligning with the trends observed in Fig. \ref{fig:avcompo}. 
Furthermore, as seen in Fig. \ref{fig:onset}, we find that the degree of supersaturation has no significant effect, while increasing the polydispersity leads to a more pronounced shift of the distributions. 



In conclusion, we studied the nucleation of polydisperse hard spheres, and demonstrated that the polydispersity has a significant effect on the nucleation rate. 
Even when mapped to the effective packing fraction, the nucleation rate of polydisperse hard spheres deviates from the trend of monodisperse hard spheres.
Furthermore, we showed that nucleation tends to originate in regions with on average more larger-sized particles, indicating that such regions act as precursors for nucleation.





\textit{Acknowledgments}---L.F. and M.d.J. acknowledge funding from the Vidi research program with project number VI.VIDI.192.102 which is financed by the Dutch Research Council (NWO). 
A.C. and F.S. acknowledge funding from the Agence Nationale de la Recherche (ANR), grant ANR-21-CE30-0051. 



\bibliography{paper}

\end{document}